\documentclass{aastex}
\usepackage{graphicx}
\usepackage{color}
\usepackage{multirow}
\usepackage{amsmath}
\begin{document}	
\title{Hemispheric Preference and Cyclic Variation of Solar Filament Chirality from 2000 to 2016}
\author{Soumitra Hazra, Sushant S. Mahajan, William Keith Douglas Jr.  \& Petrus C. H. Martens }
\affil{Department of Physics \& Astronomy, Georgia State University, Atlanta, USA 30303}
 \email{shazra@gsu.edu}

\begin{abstract}
It is well known that solar filaments are features in the solar atmosphere which show a hemispheric preference in their chirality. The hemispheric preference is such that the dextral chirality dominates in the northern hemisphere while the sinistral chirality dominates in the southern. Determining the strength and cyclic variation of the degree of this hemispheric preference however is challenging and tedious and thus needs to be automated. In this paper, we follow Dr. Pietro Bernasconi's algorithm \cite{bern05} to detect filament chirality with two parallel channels of application. The algorithm is applied to H-alpha images with the ''Advanced Automated Filament Detection and Characterization Code'' (AAFDCC) \cite{bern05} and the full algorithm (including the detection of filaments and tracking) is explained to the human observer who determines the chirality of the solar filament. We have conducted this exercise on data during the month of August from years 2000 to 2016 and we found that 83\% of our visually determined filaments follow the hemispheric chirality preference, while 58 \% of automatically determined filaments follow it. Our visually compiled results have over 90\% agreement with those of Pevtsov \textit{et al.} (2003), yet the visually determined chiralities of filaments disagree with automated determinations significantly. We find that the hemispheric preference
remained the same between solar cycles 23 and 24 but the preference is very difficult to determine during the solar minimum of 2008-2010 primarily due to the absence of filaments. 
\end{abstract}

\section{Introduction}

Solar filaments are features in the Sun's atmosphere in which a large mass of cold (compared to its surroundings) and dense plasma is suspended along magnetic field lines. Filaments can be found near active regions in which case they are called active region filaments or they can be away from active regions in relatively quiet regions in which case they are dubbed quiescent filaments. Active region filaments typically have short sizes as well as lifetimes and they are surrounded by a lot of activity. Quiescent filaments on the other hand can be longer than the radius of the Sun and can last for a few days or even a month. Various studies have shown that some solar filaments were precursors of large coronal mass ejections proposing that a filament can become unstable due to magnetic instabilities or other unknown reasons and erupt causing a huge explosion \cite{gopa20, chen08}. If such a coronal mass ejection heads towards the Earth, it causes geomagnetic storms if the direction of magnetic field in the ejected material is anti-parallel to that of the Earth's own magnetic field. This orientation of magnetic field in the ejected material may or may not depend on the magnetic field orientation in its parent filament i.e. it's chirality.

The chirality of solar filaments is a concept that was introduced by Sara Martin and her co-workers in a series of papers reviewed by Martin (1998). They showed that the spine of a solar filament lies high above a polarity inversion line on the solar surface with a strong axial magnetic field along its spine. From the point of view of a person on the positive side of the polarity inversion line, if the axial magnetic field in the filament is directed towards one's right (left), the filaments are labeled to be dextral (sinistral) in chirality. Zirker \textit{et al.} (1997) showed that dextral filaments are associated with left skewed arcades while sinistral filaments are associated with right-skewed arcades and Martens (2002) showed that this is consistent with the filament formation model of Martens \& Zwaan (2001).
 Some pieces break off from the spine of a filament to connect with patches of minority polarity flux on the solar surface on either sides of the polarity inversion line. These pieces which look like the legs of a caterpillar are called barbs and they too appear as dark as filaments in H-alpha images. Martin \textit{et al.} (1994) found that the angle at which barbs meet the spine of a filament in H-alpha images can be used to determine the direction of magnetic field in its spine and its chirality. This revelation enabled Martin \textit{et al.} (1994) to perform a systematic study of the chirality of solar filaments in which they saw a hemispheric preference for quiescent filaments but not for active region filaments. Although hemispheric helicity rule for solar prominences had been discovered as early as 1983 \cite{leroy83, leroy89}; Martin \textit{et al.} (1994) was the first to perform a statistical study on the chirality of solar filaments. They found that $100\%$ of the quiescent filaments in the northern hemisphere were dextral while $72\%$ of those in the southern hemisphere were sinistral. Almost a decade later, a visual analysis by Pevtsov \textit{et al.} (2003) found that not only $80\%$ of quiescent filaments, but $75\%$ of active region filaments also follow the same hemispheric preference of chirality. Bernasconi \textit{et al.} (2005) developed an "Advanced Automated Filament Detection and Characterization Code (AAFDCC)" and found that 68 \% of all filaments follow the hemispheric preference. The difference between the results obtained by Martin \textit{et al.} (1994), Pevtsov \textit{et al.} (2003) as well as Bernasconi \textit{et al.} (2005) is possibly due to their way of determining the filament chirality.

Although all three groups determined the filament chirality using barb orientations in H-alpha images, Martin and her coworkers believed that in a single filament only one type of barb (either dextral or sinistral) can be present. However, Pevtsov \textit{et al.} (2003) and Bernasconi \textit{et al.} (2005) determined the filament chirality without any such assumption so that both types of barbs can be present in a single filament. Both Pevtsov \textit{et al.} (2003) and Bernasconi \textit{et al.} (2005) considered the presence of a structure along the filament boundary extended away from the spine of the filament as a barb. However, Martin (2015) argued that a fine thread-like structure inside a labeled barb may consist of a series of thread ends with higher density near the footpoints; while the upper portions of these threads are invisible between dense feet and the spine (see Fig. 1) i.e. one can classify the extended structure from the side of the main filament body as a barb only if the extended structure is aligned with the underlying chromospheric fibrils. Thus, a careful consideration of barb structure is necessary to determine the chirality of solar filaments. Chen \textit{et al.} (2014) argued that Martin's procedure for determining filament chirality by barb orientation from H-alpha images is only valid for filaments which are supported by a flux rope (inverse polarity filament). They claim that Martin's rule of determining the chirality is misleading for normal polarity filaments which are supported by a sheared arcade. In a recent study, Ouyang \textit{et al.} (2017) used a new method for determining the filament chirality based on filament drainage and found that $91.6\%$ of their sample filaments during $2010-2015$ followed the hemispheric preference.

\begin{figure*}[!htb]
	\centering
	\begin{tabular}{cc}
		\includegraphics*[width=1.0\linewidth]{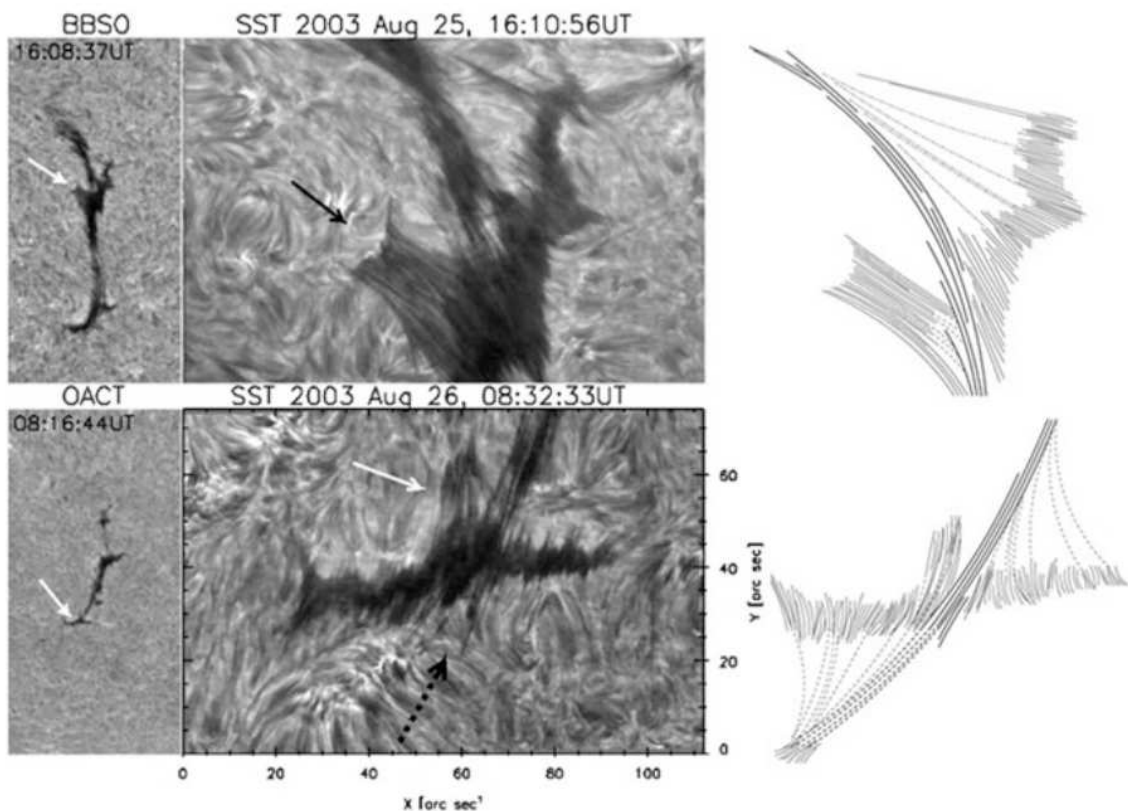} &
	\end{tabular}
	\caption{\footnotesize{This figure is adapted from Martin (2015). The white arrows in the Big Bear Solar Observatory (BBSO; top)  and the Catania Astrophysical Observatory (OACT; bottom)  images on the left point to part of the filaments observed with greater magnification and resolution at the Swedish Solar Telescope (SST) in the figure in between. On the right is a schematic drawing in which the solid lines represent observed parts of barb
		threads and dashed lines represent the invisible parts inferred by looking at the fine structure of the threads.  Illustration by Y Lin.}}
	\label{fig1}
\end{figure*}

Previous studies have reported that there is a one-to-one correspondence between the sign of magnetic helicity and filament chirality \cite{mack97,aula98, rust99}. It is often believed that helicity of an active region arises as a result of buffeting by helical turbulence during the rise of flux tube through the convection zone (Longcope \textit{et al.} 1998; Choudhuri \textit{et al.} 2004); thus one can use the sign of helicity or filament chirality as a tool to study the flux emergence process inside the solar convection zone. The behaviour of the hemispheric preference is of major importance for at least two reasons: on one hand, it is related to the helicity  and on the other hand, it allows for possible prediction of the orientation of the magnetic field when the associated cloud hits the Earth's magnetic system, a parameter of paramount importance for terrestrial (magnetic) storms.  Recent helioseismology results also confirmed the hemispheric helicity rule in local quiet \cite{kom07} and active \cite{kom14} regions on the solar surface. In this paper, we study hemispheric chirality preferences of solar filaments for a period spanning over solar cycles 23 and 24 (up to August, 2016). First, we  determine the chirality of solar filaments during this period utilizing the fully automated filament detection and characterization code (AAFDCC; Bernasconi \textit{et al.} 2005). We further visually determine the hemispheric chirality of solar filaments appearing during the month of August each year and compare it with the corresponding automated determined chirality. We provide details about the data used for this study in section 2 followed by a discussion of our results in section 3. Finally, we present the conclusions of our study in the last section.

\section{Data Analysis}	

\subsection{Data Selection}															
Our data sources in this study were the full disk H-alpha images from Big Bear Solar Observatory (BBSO) and Kanzelhoehe Observatory.															
Both provide full disk H-alpha images in FITS format from July 2000 onwards. These full disk images have a resolution less than 1 arcsec/pixel with slightly varying solar radius depending on the time of day and year. We primarily used BBSO images and reverted to Kanzelhoehe when BBSO images weren't available for that particular day. Out of the various image products, we used the daily high contrast full disk H-alpha images corrected for limb darkening from the ftp archive of Big Bear Solar Observatory (ftp://bbso.njit.edu/pub/archive/) from 2000 to 2016.

 As it is labour intensive to determine the chirality of each filament manually, we analyzed filaments from images taken every other day in August of each year. The month of August was chosen for our analysis because this is usually the least cloudy month of the year at BBSO. Our sampled data set consists of 3480 filaments. Our sampled data set includes daily observations of filaments which means filaments that last a few days are observed multiple times. Some previous studies have indicated that the determination of filament chirality at a single epoch could be very misleading partly due to geometric effects \cite{yeat07}. Our sampled data set is publicly available
in the Harvard Dataverse repository \cite{haz18}.

\subsection{Methods}

We ran the Advanced Automated solar Filament Detection and Characterization Code (AAFDCC; Bernasconi \textit{et al.} 2005) on our selected full disk H-alpha images. This code first eliminates the sunspot structures from full disk images by setting a threshold. Generally pixels associated with sunspot structures have an intensity less than -3000. After eliminating the spot structures, the code generates a filament mask in which all pixels associated with filaments within a heliocentric angle of $60^{\circ}$ have been labelled as 1 and the rest as zero. This code identifies the filament structure by setting a filament detection threshold 600. The code also uses a second filament threshold value of -500 to identify the filaments properly by removing spurious pixels and small areas which are not actually filaments.  Negative intensity value in a pixel generally corresponds to dark pixel and this threshold values for filament identification are chosen by trial and error method. After filament identification, the code determines the filament boundary and spine of the filament by a principal curve algorithm \cite{keg20}. Next the code determines the barb by calculating the distance of each pixel in the boundary array from the spine. Boundary array is an array which stores the cartesian coordinate values of each pixel along the filament boundary. Finally, the code identifies the barb as dextral and sinistral only when the barb makes an angle which is within +5/-5 degree with the spine. In summary, this code determines the position of filaments and their chirality automatically. For calibration, we determined the chirality of solar filaments visually and compared our results with the automatic computer generated results. We determined the chirality of the sampled filaments following the same definition as Bernasconi \textit{et al.} (2005): if $R$ and $L$ denote the number of right bearing and left bearing barbs respectively then $R-L \geq 2 $ corresponds to a left-handed filament, $R -L \leq −2$ corresponds to a right-handed filament, and $|R -L| < 2$ corresponds to an undetermined chirality. Here, a right-handed filament corresponds to sinistral chirality and respectively a left-handed filament corresponds to dextral chirality. In visual determination, the left bearing and right bearing barbs of a filament were identified by looking at their pixel locations and orientation in the images.  Martin (2015) points out that the underlying fine threads give the true direction of the magnetic field inside the barb.  For this study, we assume that barb orientation reflects the underlying thread orientations, even though this may not always be the case. The chirality of some filaments was also determined visually by people other than the first author to minimize the subjectivity in the visual determination. We also performed a statistical {\it t}-test analysis following the prescription of Yeates \textit{et al} (2007), to minimize the subjectivity in determining the chirality of filaments. In this procedure, suppose $N_{right}$ is the number of right-bearing barbs in a single filament, then the number of left-bearing barbs is $N_{left}= N - N_{right}$ where $N$ is the total number of classified barbs in a single filament. If we assume that the number of right-bearing barbs follows a binomial distribution with parameters (N, p), then this procedure is based on the following statistic:
\begin{equation}
 {\it t}= \frac{N_{right}-Np}{\sqrt{Np(1-p)}}
\end{equation}
where we assume $p=0.5$. Now, if T is some chosen threshold, then ${\it t} > T$ corresponds to a dextral chirality filament, ${\it t} < -T$ corresponds to a sinistral chirality filament and $|{\it t}| \leq T$ corresponds to an undetermined chirality filament. We take the value of the chosen threshold $T$ as a threshold ratio $N_{right}/N= \frac{3}{(4 /\sqrt{N})} + 1/2$. This threshold like the prescription from Bernasconi et al. (2005) only determines the chirality of filaments with two or more barbs. This is why we have chosen to eliminate filaments with less than two barbs from our analysis.

\section{The Hemispheric Chirality Rule and It's Cyclic Behaviour}
In our visually sampled data set, 1722 filaments out of 3480 appear in the northern hemisphere and the rest are in the southern hemisphere. We first use Bernasconi's ($R-L$) criteria for the determination of chirality. We find that most of the filaments on the solar disk have undetermined chirality. We were able to determine specific chirality only for 388 filaments in the northern hemisphere and 374 filaments in the southern hemisphere. 85\% of filaments (330 out of 388) in the northern hemisphere have dextral chirality; and 81\% of filaments (303 out of 374) in the southern hemisphere have sinistral chirality. Thus, together 83\% of our visually sampled determined filaments follow the hemispheric chirality preference. We can also use the {\it t}-statistic analysis to determine the chirality of our sampled filaments. In that case, we find that 84\% of our visually sampled filaments follow the  same hemispheric preference (see Table 1 for a detailed comparison of hemispheric preference between two different definitions of chirality determination).
\begin{table}[!htb]
	\begin{center}
		\begin{tabular}{ |c|c|c|c| }
			\hline
			\multicolumn{2}{|c|}{} & \multicolumn{2}{|c|}{Determination of Chirality} \\
			\multicolumn{2}{|c|}{} & Bernasconi Definition& {\it t}-test Statistics \\
			\hline
			\multirow{4}{4em}{AAFDCC Code} & Dextral Filaments in the Northern Hemisphere & 63 \% & 57 \% \\
			& Sinistral Filments in the Northern Hemisphere & 37 \% &43 \% \\
			& Dextral Filaments in the Southern Hemisphere  & 47 \% & 43 \% \\
			&Sinistral Filaments in the Southern Hemisphere &  53 \% & 57 \% \\
			\hline
			\multirow{4}{4em}{Visual} & Dextral Filaments in the Northern Hemisphere & 85 \% & 83 \% \\
			& Sinistral Filments in the Northern Hemisphere & 15 \% & 17 \% \\
			& Dextral Filaments in the Southern Hemisphere  & 19 \% & 15 \% \\
			&Sinistral Filaments in the Southern Hemisphere &  81 \% & 85 \% \\
			\hline
		\end{tabular}
		\caption{Comparision between hemispheric preference of filament chirality obtained from two different methods of chirality determination.}
	\end{center}
\end{table}

\begin{figure*}[h]
	\centering
	\begin{tabular}{cc}
		\includegraphics*[width=1.0\linewidth]{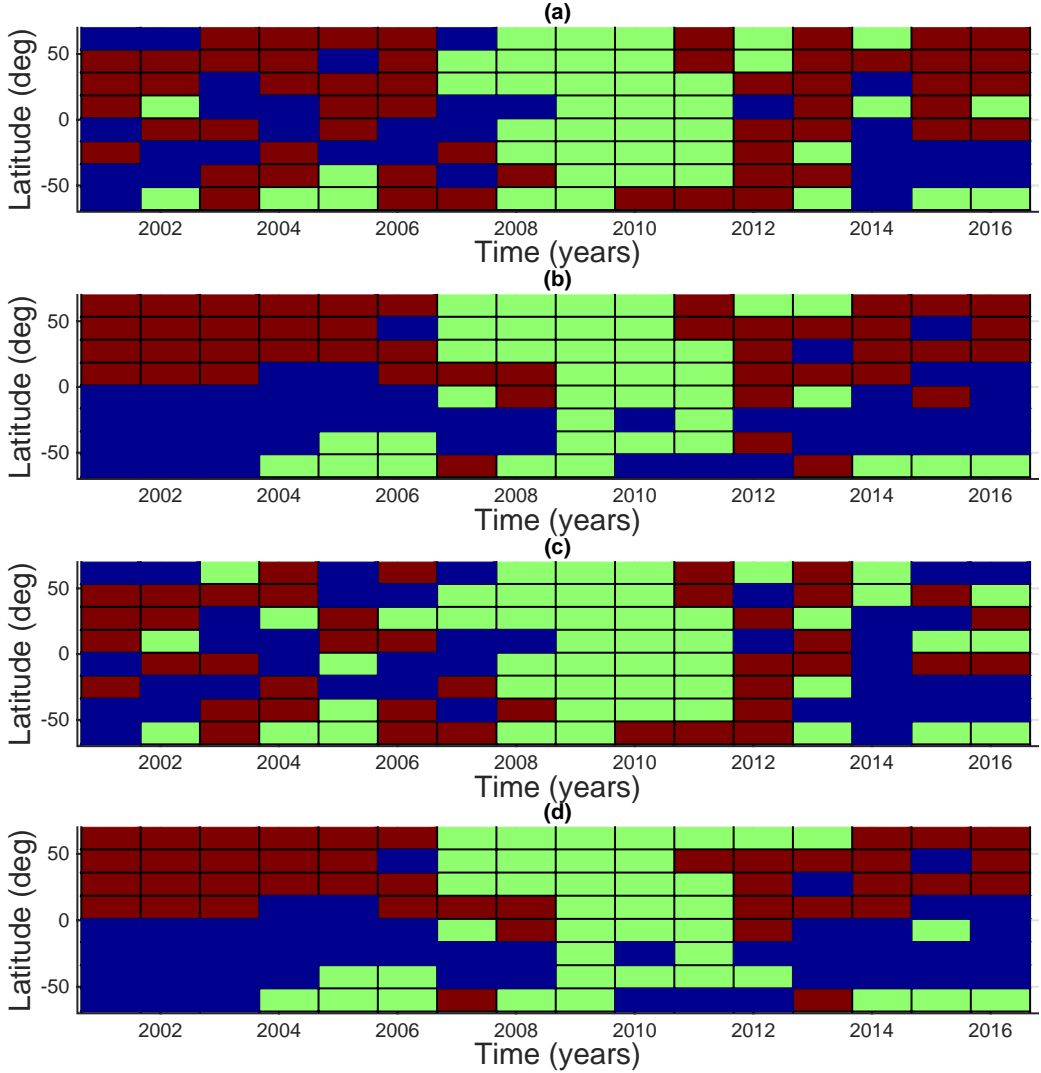} &
	\end{tabular}
	\caption{\footnotesize{Figs. (a) and (b) show the latitude-time plot of the sign of average chirality of all filaments corresponding to each pixel in the plot obtained from the AAFDCC and our visual determination respectively, when we determine the chirality for both methods following the Bernasconi \textit{et al.} (2005) prescription.  Figs. (c) and (d) show the same, when we determine the chirality for both methods following the {\it t}-test statistic. Red indicates dextral chirality, blue sinistral, and green denotes no filament with determined chirality. }}
	\label{fig2}
\end{figure*}
\begin{figure*}[!h]
	\centering
	\begin{tabular}{cc}
		\includegraphics*[width=1.0\linewidth]{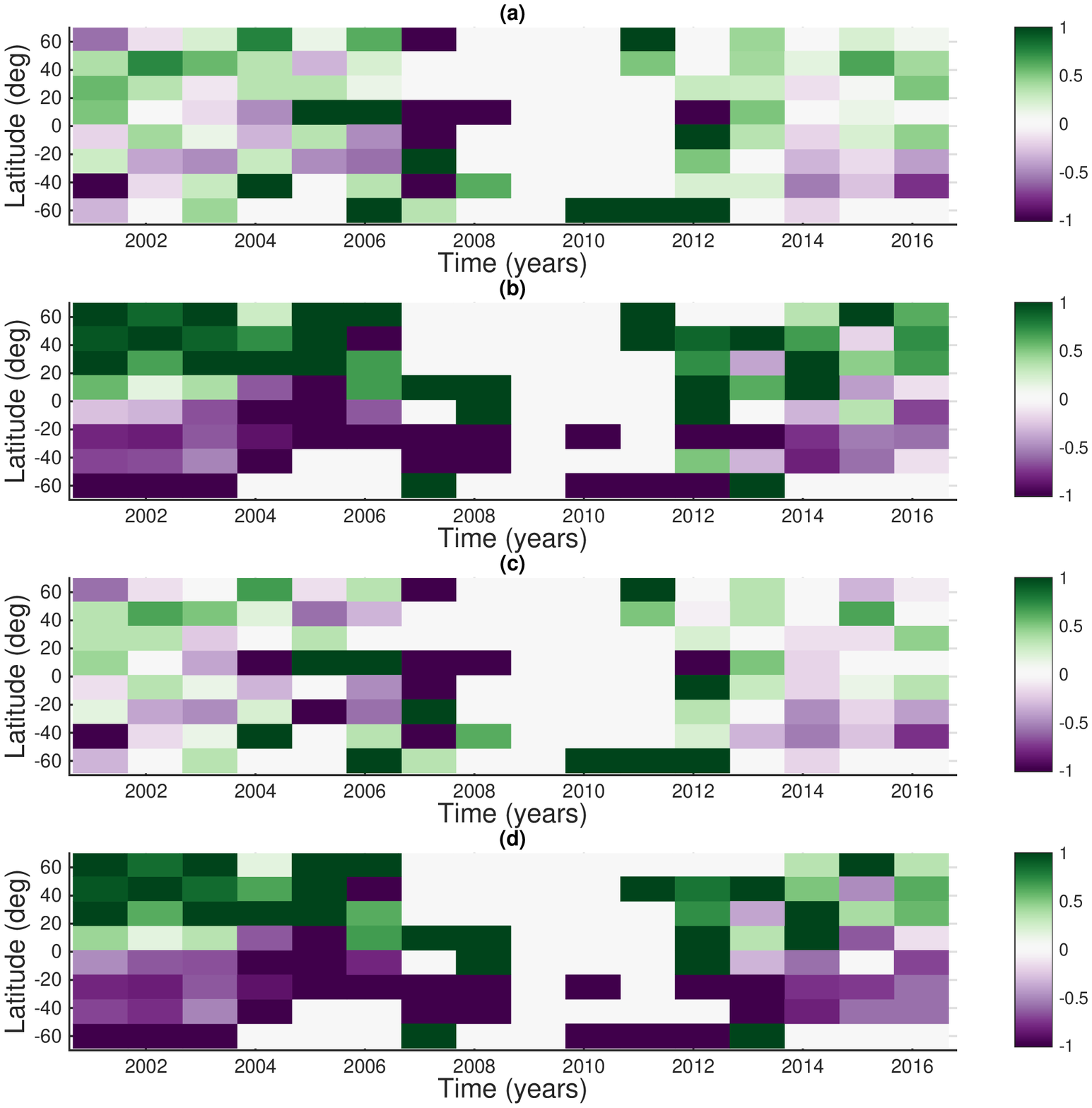} &
	\end{tabular}
	\caption{\footnotesize{Figs. (a) and (b) show the latitude-time plot of the average of fractional chirality of all filaments corresponding to each pixel obtained from the AAFDCC and visual determination respectively, when we determine the chirality for both methods following the Bernasconi \textit{et al.} (2005) prescription.  Figs. (c) and (d) show the same, when we determine the chirality for both methods following the {\it t}-test statistic. This plot shows the latitudinal variation of the strength of hemispheric preference. Zero value pixel in the plot denotes absence of filaments with determined chirality.}}
	\label{fig3}
\end{figure*}

 AAFDCC on the other hand, determines the number of left and right-bearing barbs automatically, and we ascertain the chirality of our sampled filaments using both the Bernasconi \textit{et al.} (2005) definition and the {\it t}-test statistic. The automated code was able to determine the chirality of 427 filaments in the northern hemisphere and 325 filaments in the southern hemisphere, using Bernasconi \textit{et al.} (2005) definition. It found 63\% of filaments (270 out of 427) in the northern hemisphere with dextral chirality, and 53\% of filaments (174 out of 325) in the southern hemisphere with sinistral chirality. This automated result for this limited data set also follows the preference of hemispheric chirality, albeit rather weakly. Using {\it t}-test statistic to determine the chirality, we find that 57\% of filaments in the automated results follow the same hemispheric preference (see Table 1 for a detailed comparison). 
 
Figure 2 shows the latitude-time plot of the sign of mean chirality obtained from both AAFDCC and visual determination. A closer inspection of Figure 2 (a and c) reveals that the mean chirality in both hemispheres obtained by the AAFDCC code becomes ambiguous near the solar minimum (2008-2009) and at the beginning of the solar cycle (2010-2011) i.e. the hemispheric chirality preference appears as if it is not properly established. Visual chirality determination also shows that the hemispheric preference is uncertain during the solar minimum (2008-2009) and the beginning phase of the solar cycle (2010-2011) with a distinction: it is also ambiguous at the end of cycle 23 (2007)  (Fig.~2 (b) and (d)). Martens \textit{et al.} (2014) also determined the chirality of all filaments for solar cycle 23 and the initial phase of cycle 24 (2000-2012) using the same AAFDCC code as us and found that most of the filaments in cycle 23 follow the hemispheric chirality rule but that the preference disappears during the solar minimum (2008-2009) and has a hard time re-establishing itself during the intial phase of cycle 24. The possibility of hemispheric rule violation at solar minimum was also indicated using numerical simulations by Yeates \& Mackay (2012). 
\begin{figure*}[!htb]
	\centering
	\begin{tabular}{cc}
		\includegraphics*[width=1.0\linewidth]{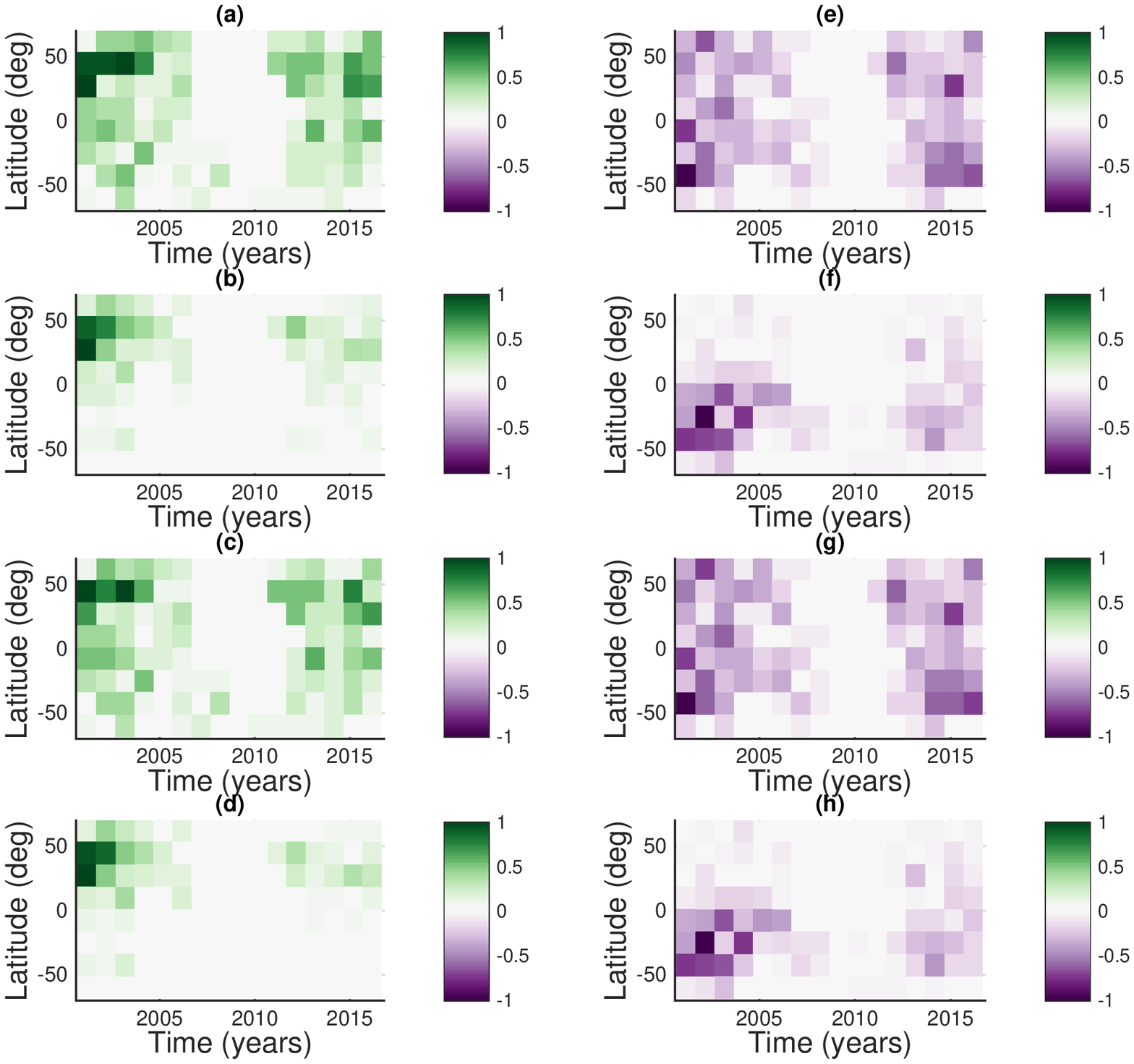} &
	\end{tabular}
	\caption{\footnotesize{Figs. (a) and (b) show the latitude-time plot of the normalized mean dextral chirality of all dextral filaments corresponding to each pixel obtained from the AAFDCC and visual determination respectively, when we determine the chirality following Bernasconi \textit{et al.} (2005) prescription; Figs. (c) and (d) represent the same when we follow the t-test statistic.  Figs. (e) and (f) show the latitude-time plot of the normalized mean sinistral chirality of all sinistral filaments corresponding to each pixel obtained from the AAFDCC and visual determination respectively, when we determine the chirality  following the Bernasconi \textit{et al.} (2005) prescription and  Figs. (g) and (h) represents the same when we follow the t-test statistic.} }
	\label{fig4}
\end{figure*}
\begin{figure*}[!htb]
	\centering
	\begin{tabular}{cc}
		\includegraphics*[width=1.0\linewidth]{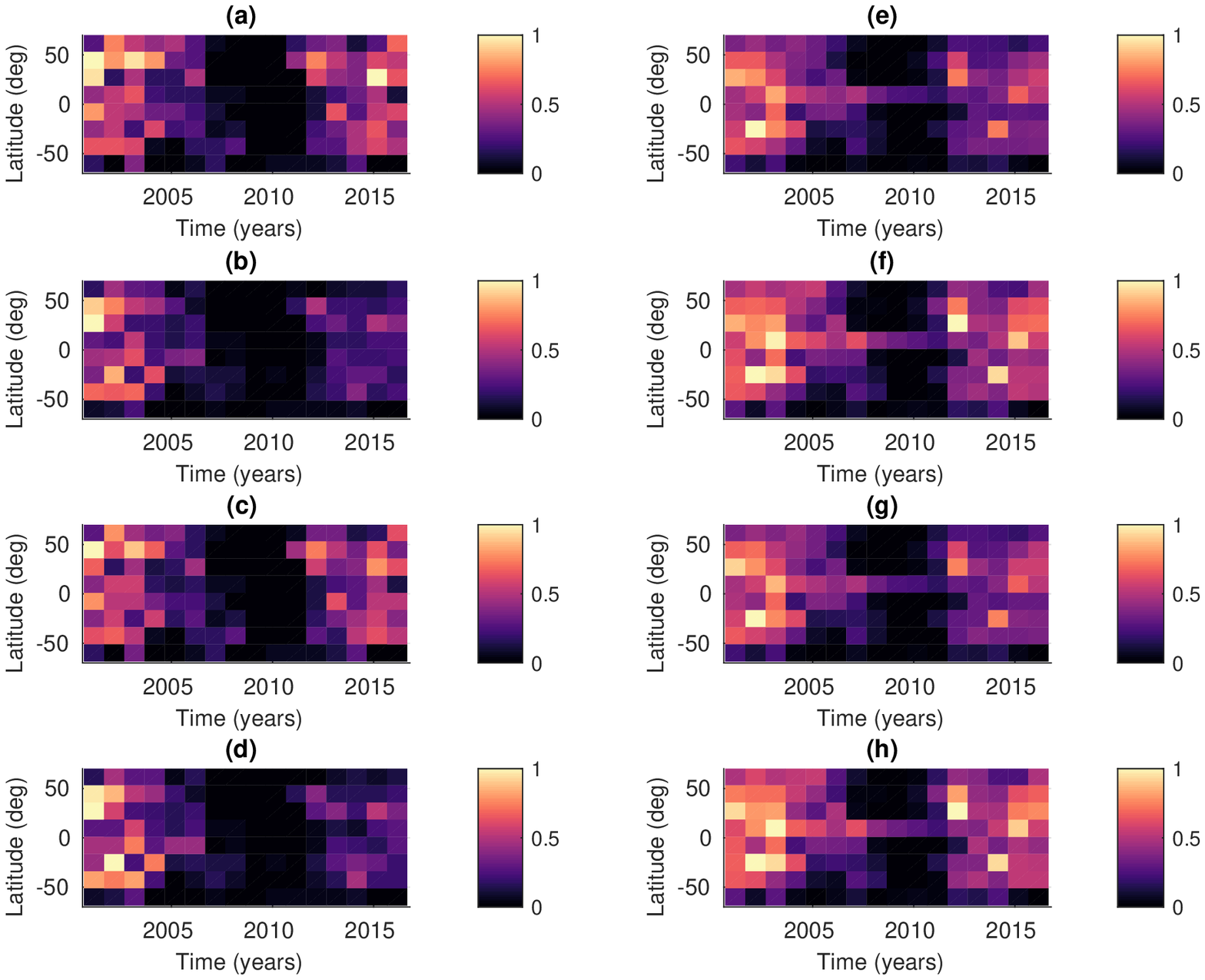} &
	\end{tabular}
	\caption{\footnotesize{Figs. (a) and (b) show the latitude-time plot of the normalized mean of the number of determined filaments corresponding to each pixel obtained from the AAFDCC and visual determination respectively, when we determine the chirality following Bernasconi \textit{et al.} (2005) prescription and Figs. (c) and (d) represent the same when we follow the t-test statistic. Figs. (e) and (f) show the latitude-time plot of the normalized mean of the number of undetermined filaments corresponding to each pixel obtained from the AAFDCC and visual determination respectively, when we determine the chirality following Bernasconi \textit{et al.} (2005) prescription and Figs. (g) and (h) represent the same when we follow the t-test statistic.}}
	\label{fig5}
\end{figure*}

We have calculated the fractional chirality ($(N_{dex}-N_{sin})/(N_{dex}+N_{sin})$) for each filament where $N_{dex}$ and $N_{sin}$ are the numbers of left and right-bearing barbs in a single filament respectively. The latitude-time plot of fractional chirality (Fig.~3) again indicates that hemispheric chirality preference is uncertain at the time of solar minimum. The latitude-time plot of dextral and sinistral chirality filaments (Fig.~4) shows that both are more prevalent at the peak phase of the solar cycle and appear in large numbers at mid to low latitude in both hemispheres. However, we did not find any significant pattern in the hemispheric strength variation with latitude (see Fig.~3). The latitude-time plot of the normalized number of determined and undetermined filaments (Fig. 5) indicates that the number of both determined and undetermined filaments are maximum at the peak phase of the solar cycle and appear more at mid latitudes in both hemispheres. At the end and at the initial phase of the solar cycle, most of the filaments have undetermined chirality.

\begin{figure*}[!htb]
	\centering
	\begin{tabular}{cc}
		\includegraphics*[width=1.0\linewidth]{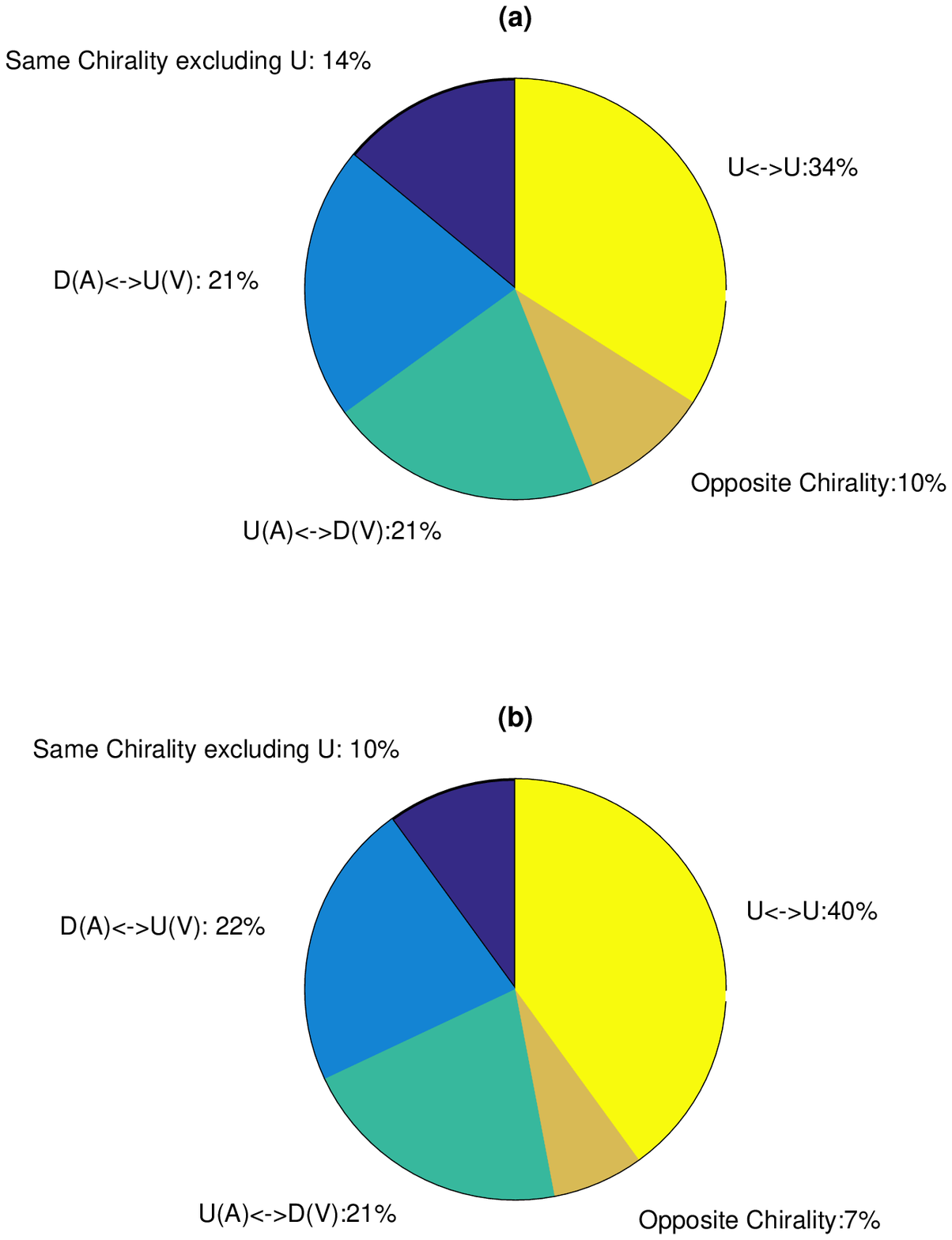} \\
	
	\end{tabular}
	\caption{\footnotesize{Pie charts (a) and (b) show the filament-by-filament comparison between automatic and visual determination, when we determine the chirality for both methods following Bernasconi \textit{et al.} (2005) prescription and t-test statistic respectively. From this chart, it is clear that only few filaments match chirality between methods. ''D'' and ''U'' symbols denote determined and undetermined respectively. ''A'' and ''V'' symbols inside the bracket correspond to automatic and visual respectively.}}
	\label{fig6}
\end{figure*}

\begin{figure*}[!htb]
	\centering
	\begin{tabular}{cc}
		\includegraphics*[width=1.0\linewidth]{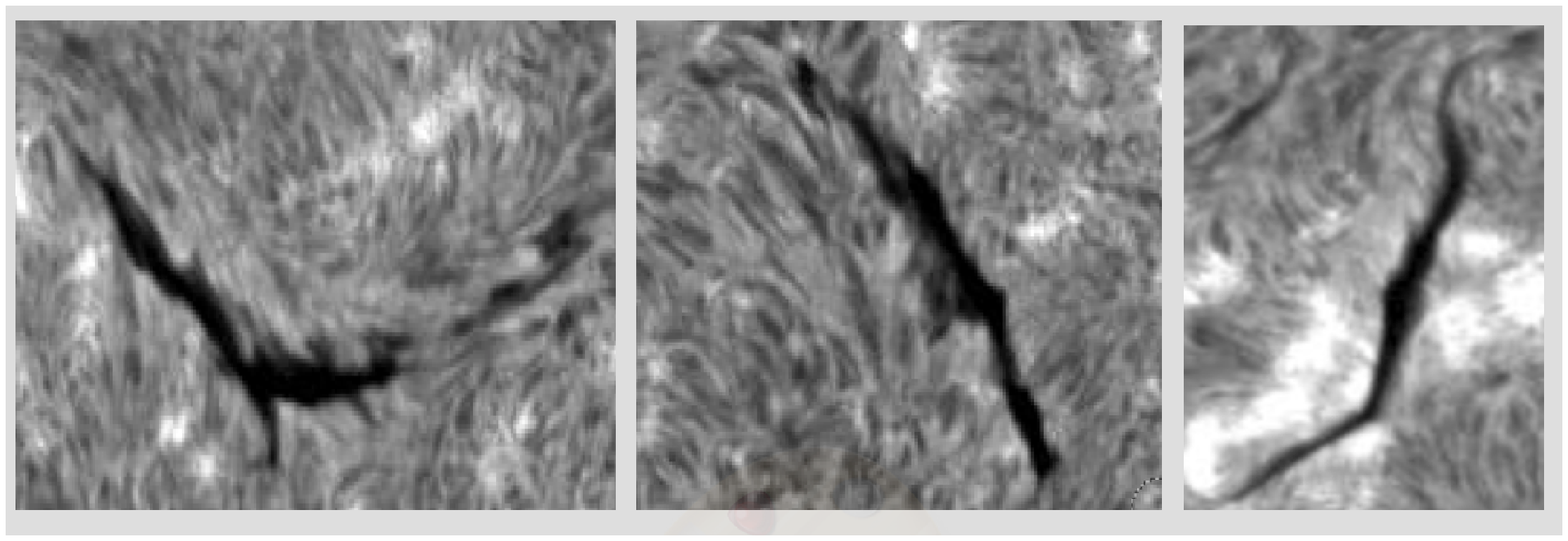} &
	\end{tabular}
	\caption{ \footnotesize{Left plot is an example of a BBSO H-$\alpha$ filament where AAFDCC and visual chirality determination agree. Both AAFDCC and visual determination detect this filament as dextral. Middle plot is an example of a BBSO H-$\alpha$ filament where AAFDCC and visual chirality determination don't agree. While AAFDCC detects this filament as sinistral, visual determination detects it as dextral. Right plot is an example of a BBSO H-$\alpha$ filament where both AAFDCC and visual method suggests that the filament has no determined chirality. The difference between the human observer and AAFDCC arises probably due to the choice of barbs or a low-resolution image.}}
	\label{fig6}
\end{figure*}

We also compared our visual results with the automated results filament-by-filament. There are 1351 filaments after removing the filaments which have less than two barbs in any one of the two lists.  From Figure 6(a) we found that there is only a 14 \% match in the chirality of solar filaments with determined chirality between the visual and automated lists. Another 34 \% filaments match which are undetermined in both lists. 10 \% have opposite chirality between both methods and 21 \% of filaments changed their chirality from determined in the automatically compiled list to undetermined in the visually compiled list and the same percentage of filaments also changed their chirality from undetermined in the automatically compiled list to determined in the visually compiled list. Using the {\it t}-test statistic, we found only a 10 \% match between the visual and automated chirality list (see Figure 6(b) for a detailed comparison). To check the sanity of our visual determinations, we compared our results with Pevtsov \textit{et al.} (2003), who also determined the chirality of some solar filaments visually and found that 30 filaments were present in both lists. Out of these 30 filaments, the chirality of 27 filaments matched; and only three filaments changed from determined chirality in Pevtsov's list to undetermined in our list. This indicates that there is a 90\% match between our visually determined chirality and Pevtsov's list albeit for a small sample. However, we note that if a filament has undetermined chirality with some method, it does not mean that the method disagrees with all other methods. If we focuse only on filaments with chirality determined by both methods, the disagreement according to Fig. 6a (with the $|R-L|$ criterion) is only 10\% (of all considered filaments), while agreement is 14\%.

In another test, we provided 32 filaments to a non solar physics expert for determination of chirality. While we assigned chirality to 12 filaments, he only assigned a definite chirality to seven. His seven determined filaments agree with our visual determination, while only three agree with the automated determination. It seems as if non-experts are more reluctant to assign chirality than solar physicists.

Figure 7 shows three examples of low-resolution BBSO H$\alpha$ filaments where, for left image visual determination agrees with the automatic, for middle filament image, it doesn't and in the right filament, both automated and visual method suggests that the filament has no determined chirality. We confirm that visual chirality determination is susceptible to error due to person-to-person differences in judging the number of barbs and their direction. However, the human eye can detect finer and more subtle threads, while the automated code is only able to determine the chirality of dark and visible barbs. We also note that according to our definition of chirality determination, both types of barbs (left-bearing and right bearing) can co-exist in a single filament, while Martin (2015) argues that only one type of barb (either left-bearing or right-bearing) can exist in a single filament. It is difficult to know exactly how many filaments in visual determination are assigned the wrong chirality owing to the fact that some of the barbs visible in the BBSO H-alpha images could be aligned oppositely to the underlying chromospheric fibril structures. Our determination of clear and visible barbs (both automated and manual) may consist of a series of thread ends with high density near the footpoints, while the upper portion of these threads is invisible due to less density. This argument needs further investigation with higher resolution images.

 In short, we found that the hemispheric chirality preference holds well from one cycle to another (Figs. 2 and 3) although the hemispheric preference can not be determined at the time of solar minimum. This may be due to the fact that there are very few filaments with more than one barb during this phase of the solar cycle. Note that we have picked only the filaments which have two or more barbs for this analysis, which means we had to ignore a large number of filaments with one or no barb.

\section{Conclusions}										
We performed a statistical study of filament chirality from 2000 to 2016. This period encompasses almost all of solar cycle 23 and some part of solar cycle 24.  The number of left and right-bearing barbs was determined both automatically (using AAFDCC) and visually (manually). The chirality of each filament was determined using the definition of chirality prescribed by Bernasconi \textit{et al.} (2005). We also performed a statistical t-test analysis to quantify the uncertainty in the determination of chirality of each filament. Our main conclusions are summarized below:

First,  about 83\% of our visually determined filaments follow the hemispheric preference, while 58\% of automatically determined filaments do the same. Our visual determination results do not agree at all with the automated determination results in a filament by filament comparison, even though they show similar overall hemispheric preference. This is probably because the human eye is capable of detecting finer and subtle threads which sometimes could even appear to be slightly disconnected from the spine of the filament, while the automated code is only designed to detect the dark, broad and visible barbs which are connected to the spine of the filament. 

Second, the overall hemispheric preference of filament chirality, however, does not reverse from one cycle to another regardless of the method used to determine chirality and the hemispheric preference is ambiguous during the solar minimum due to the lack of filaments with more than one barbs.	

Third, the latitude-time plot of dextral and sinistral chirality filaments indicate no significant variation in hemispheric strength with latitude. It also shows that both kind of filaments appear more at the peak phase of the solar cycle and in large numbers at the mid to low latitude in both hemisphere.

Fourth, the latitude-time plot of determined and undetermined chirality filaments indicate that most of the filaments have undetermined chirality at the end and initial phase of the solar cycle.	 It also shows that both determined and undetermined filaments appear more in mid to low latitude in both hemisphere and at the peak phase of the solar cycle

Finally, we do not believe that we have a choice between the automated and visual method. Given the enormous amount of solar data, not just for filaments, we must build calibrated and reliable automated methods for analyzing images. What this paper clearly underscores is that the automated methods must be thoroughly vetted before let loose. We believe the automated method could (and must) be improved.  AAFDCC works fine for detecting, outlining, and now also tracking filaments, as well as detecting barbs.  We are looking now at fixes to get the barb inclination right, e.g. calculating the Tamura texture directionality for the detected barbs (see Ahmadzadeh \textit{et al.} 2017). However, there is a more fundamental issue here, as commented on in the introduction, in that the barb inclination not always produces the correct axial magnetic field direction, and hence chirality. We have started a project to set up a pipeline to detect filament chirality directly from GBO vector magnetograms. A successful result will allow us to detect the chirality of filaments in active regions much more easily then, which is relevant for the prediction of large flares and geo-effective eruptions.

\acknowledgements{We thank the anonymous referee for valuable comments and suggestions. We also thank Sara Martin, Alex Pevtsov, Pietro Bernasconi, Duncan Mackay and Anthony Yeates for reading the manuscript and providing valuable suggestions. This work is supported by NASA grant NNX15AQ61G, Formation, Evolution and Eruption of Solar Filaments for a Full Cycle: Simulation Verified by Observations.}

\end{document}